\documentclass[%
 aip,
 amsmath,amssymb,
 reprint,%
]{revtex4-1}
\pdfoutput=1
\usepackage[latin9]{inputenc}
\usepackage[english]{babel}
\usepackage{graphicx}
\usepackage{bm}
\usepackage{color,soul}
\usepackage{dcolumn} 
\usepackage{amsfonts}
\usepackage{natbib}
\usepackage{upgreek}
\usepackage{amssymb}
\usepackage{amsmath}
\usepackage{physics}
\usepackage{graphicx}
\usepackage{lipsum}
\usepackage{float}
\usepackage[dvipsnames]{xcolor}
\usepackage{url}
\usepackage{hyperref}
\usepackage[normalem]{ulem}
\hypersetup{
    colorlinks=true,
    urlcolor=blue,
    linkcolor=blue,
    citecolor=blue}
\usepackage{upgreek}

\newcommand{\otherlabel}[2]{\protected@edef\@currentlabel{#2}\label{#1}}

\begin{document}

\title{Temperature-dependent critical spin-orbit field for orthogonal switching in antiferromagnets}

%\title{Temperature dependence of the critical field for spin-orbit orthogonal switching in antiferromagnets}

\author{R. Rama-Eiroa}
\email{ricardo.rama@ehu.eus}
\affiliation{Donostia International Physics Center, 20018 San Sebasti\'an, Spain}
\affiliation{Polymers and Advanced Materials Department: Physics, Chemistry, and Technology, University of the Basque Country, UPV/EHU, 20018 San Sebasti\'an, Spain}
\affiliation{EHU Quantum Center, University of the Basque Country, UPV/EHU, 48940 Leioa, Spain\looseness=-1}
\author{R. M. Otxoa}
\affiliation{Hitachi Cambridge Laboratory, J. J. Thomson Avenue, Cambridge CB3 0HE, United Kingdom\looseness=-1}
\affiliation{Donostia International Physics Center, 20018 San Sebasti\'an, Spain}
\author{U. Atxitia}
\email{u.atxitia@csic.es}
\affiliation{Instituto de Ciencia de Materiales de Madrid, CSIC, Cantoblanco, 28049 Madrid, Spain\looseness=-1}
%\affiliation{Fachbereich Physik, Freie Universit\"at Berlin, Arnimallee 14, 14195 Berlin, Germany}
\affiliation{Dahlem Center for Complex Quantum Systems and Fachbereich Physik,  Freie Universit\"{a}t Berlin,  14195 Berlin, Germany}

\date{\today}

\begin{abstract}
The discovery of current-induced spin-orbit torque (SOT) orthogonal reorientation, also known as orthogonal switching, of metallic Mn$_2$Au and CuMnAs has opened the door for ultrafast writing of an antiferromagnet (AFM). Phenomenological theory predicts that the minimum field necessary for SOT switching  -- critical field -- for ultrashort pulses increases inversely proportional to the pulse duration, thereby limiting the use of ultrafast stimulus as driving force for switching. We explore the possibility that by varying the working temperature the critical field reduces enabling orthogonal switching in response to ultrashort pulses. To do so, we extend previous theory to finite temperature and show that the critical field for an orthogonal switching strongly depends on temperature. We determine how the temperature dependence of the critical field varies as a function of the pulse duration. While for long pulses, the temperature dependence of the critical field is determined by the anisotropy field, for ultrashort pulses, it is determined by the characteristic frequency of the AFM. We show that the short and long pulse duration limits for the critical field can be connected by an analytical expression.  
%The temperature-dependent N\'eel order parameter switching of a simple cubic-based layered antiferromagnet under the action of staggered fields is explored. The power scaling laws that govern, through the macrospin approach, the dynamic process have been validated through atomistic spin dynamics simulations in the presence of white noise-based thermal fluctuations.The inertia-driven final state of the staggered vector in the after-pulse anisotropy-based energy landscape depends on its relative in-plane angular position with respect to the field-governed global energy minima, which is very sensitive to the thermal background under the same external conditions. To achieve a reliable picosecond-based orthogonal switching, it is necessary to use pulse durations compatible with a sufficiently damped oscillation around the easy-axis, which becomes more robust as the temperature increases, while the critical field decreases according to an Arrhenius-like law, facts which are of vital interest in the framework of real experiments excited by ultrafast excitation processes.
\end{abstract}

\maketitle

Research of electric control of magnetic order is driven by the quest to find faster and more energy efficient platforms to store and manipulate information. Certainly, the most remarkable observation in this research field is the magnetic order reorientation upon electrical excitation in AFM. Already a number of AFM systems have been shown to exhibit electric switching: these include  CuMnAs\cite{wadley2016electrical,janda2020magneto} and Mn$_2$Au \cite{bodnar2018writing,meinert2018electrical, godinho2018electrically,Zhou2018} via SOT\cite{vzelezny2014relativistic} and NiO via spin-transfer torque \cite{Chen2018,Chirac2020}. Only very recently research has started to address the important question about which conditions allow for an electric-driven ultrafast switching of an AFM \cite{olejnik2018terahertz,Kaspar2020}. While a theoretical study predicts robust ultrafast switching using picosecond electric pulses in Mn$_2$Au \cite{roy2016robust}, an experimental demonstration is missing. 
The theory suggests that depending on the pulse duration two regimes exist. For pulse duration longer than a critical value, the magnetic order is quasi-statically driven over the energy barrier. Once the barrier is overcome, the reorientation process is driven by a minimisation of the anisotropy energy. For shorter pulse duration the switching scenario changes, the energy provided to the system is insufficient to overcome the energy barrier and inertial effects are necessary to reorient the magnetic order.  In this scenario, the critical field increases inversely with the pulse duration which dramatically increase the critical field necessary for switching. A well-known method to reduce the critical field for long, quasi-static pulses is to employ the strong temperature dependence of the anisotropy to reduce the energy barrier \cite{Kryder2008}. For short pulses, though, one has to consider the combined effect of both the temperature dependence of the energy barrier and the inertial effects on the reorientation conditions. Insights about the temperature dependence of the critical field for short duration pulses could open the possibility for a ultrafast control of the electric-induced reorientation of the magnetic order in AFM.

In this work, we extend to finite temperature a previous zero temperature theory for the description of electric-induced reorientation in AFM. The temperature scaling of the magnetic parameters, that is, anisotropy, exchange,  magnetization, and Gilbert damping, is assumed to be the same as the well-known scaling laws for ferromagnets (FM). We validate these scaling laws by direct comparison between theory and computer simulations based on atomistic spin dynamics (ASD) \cite{Nowak2007,Evans2014}.
We propose an analytical expression for the temperature dependence of the critical field as a function of pulse duration. 
%can be used to capture the effect of the temperature on the magnetization dynamics \cite{Wienholdt2012,Selzer2016,Jenkins2021,Dannegger2021}.

The dynamics of the N\'eel order parameter of a two sublattice AFM, $\mathbf{l} = \left( \mathbf{m}_{\mathrm{A}} - \mathbf{m}_{\mathrm{B}} \right)/2$, where $\mathbf{m}_{\mathrm{A},\mathrm{B}}$ is sublattice magnetization, under a staggered field $H$ can be described through a macrospin model\cite{roy2016robust}.
%namely , where $\mathbf{m}_i$ denotes the macroscopic magnetization vector at the $i$-{\it th} magnetic sublattice.
Due to the strong constriction of the sublattice magnetization in the $xy$ plane, this model allows to characterize the order parameter dynamics through the azimuthal angle $\phi$ \cite{roy2016robust}. The staggered vector is defined as $\mathbf{l} \simeq \left( \cos \phi, \sin \phi, 0 \right)$. Since in compensated AFM, the magnetization $\mathbf{m}= \left( \mathbf{m}_{\mathrm{A}} +\mathbf{m}_{\mathrm{B}} \right)/2 \approx 0$, a uniform external magnetic field is inefficient for switching AFM.
In inversion-asymmetric AFM,  an electric current passing through the sample generates the so-called non-equilibrium SOT, which effectively translates into staggered effective magnetic fields acting on each magnetic sublattice \cite{vzelezny2014relativistic}. For AFM with tetragonal anisotropy and driven by such SOT, the dynamics of the azimuthal angle defining the magnetic order parameter is described by 
%%%%%%%%%%%%%%
\begin{equation}
\ddot{\phi}+\frac{\omega^2_{\mathrm{R}}}{4} \sin 4 \phi -2 \gamma \omega_{\mathrm{e}} H \cos \phi+2 \alpha \omega_{\mathrm{e}} \dot{\phi}=0,  
\label{eq:1}    
\end{equation}
%%%%%%%%%%%%%%
where the AFM exchange-based frequency is defined as $\omega_{\mathrm{e}}= \abs{5J} \gamma/\mu_{\mathrm{at}}$, where $1.447 |J|= k_{\mathrm{B}} T_{\mathrm{N}}$ \cite{Garanin1996},
where $T_{\mathrm{N}} = 1000 $ K is the N\'eel temperature.
We note that critical temperatures of around $1300$ to $1600$ K have been found in Mn$_2$Au\cite{barthem2013revealing} and $480$ K in CuMnAs\cite{Wadley2015}. 
%($J_1$ and $J_2$ inter-layer exchange coupling constants) and
$\mu_{\mathrm{at}} = 4 \mu_{\rm{B}}$ is the atomic magnetic moment.   
$\alpha$ encodes the macroscopic Gilbert damping parameter. The frequency $\omega_{\mathrm{R}}$ is defined in terms of the exchange interaction and tetragonal anisotropy ($d_{4,\|}$) frequencies, the latter being given by $\omega_{4 \parallel}=2 \gamma d_{4 \parallel}/\mu_{\mathrm{at}}$, such that $\omega_{\mathrm{R}}=\sqrt{2 \omega_{\mathrm{e}} \omega_{4 \parallel}}$. 
We use the following parameters in our study: At $T=0$ K, $\omega^0_{\mathrm{e}}= \left( 1551 \, \mathrm{T} \right) \gamma$,  $\omega^0_{4 \parallel}=\left( 164.5 \, \mathrm{mT} \right) \gamma$, and $\omega^0_{\mathrm{R}}=\sqrt{2 \omega^0_{\mathrm{e}} \omega^0_{4 \parallel}}$.

The critical field, $H_{\mathrm{crit}}$, necessary to switch or reorient the magnetic order in tetragonal AFM by $90^{\circ}$ has been estimated for long pulse duration. The last term in Eq. \eqref{eq:1}, whose coefficient is defined by the exchange frequency and damping constant, can be neglected as the inertial effects ($\ddot{\phi}$ in Eq. \eqref{eq:1}) become sufficiently  small\cite{roy2016robust}.
This reduces the condition for switching to that $\mathbf{l}$ overcomes the tetragonal anisotropy barrier, determined by the value for $d_{4,\|}$.
For $\tau_{\mathrm{p}} \rightarrow \infty$, the critical field for switching has been estimated to be $\gamma H_{\rm{crit}}= (\omega_{4 \parallel}/ 4) \left[\, \mathrm{max} \left( \sin \left( 4\phi \right)/\cos \left( \phi \right) \right) \right]$ in the interval $\phi\in (0,\pi/4)$ ($\mathrm{max} [\sin(4\phi)/\cos(\phi)]=1.088)$. 
By the numerical solution of Eq. \eqref{eq:1}, we find that this exact expression works fine for relatively small anisotropy values, e.g. $10 \, \mathrm{mT}$ used in Ref. \onlinecite{roy2016robust}. For higher anisotropy values, however we find that this expression is not exact anymore.
The numerical value $\gamma H_{\rm{crit}} \left( \tau_{\mathrm{p}} \rightarrow \infty \right)$ for high anisotropy is still of the order of $\omega_{4,\|}$ but approximately a 25\% lower than predicted. In the following  $\gamma H_{\rm{crit}} \left( \tau_{\mathrm{p}} \rightarrow \infty \right)$ is determined directly from the numerical calculation of Eq. \eqref{eq:1}.
%For calculations including the temperature dependence of the parameters, which will be discussed later on, the temperature dependence of $\gamma H_{\rm{crit}} (\tau_{\mathrm{p}} \rightarrow \infty)$ resembles closely to that of the tetragonal anisotropy
%%%%%%%%%%%%%%%%
%\begin{equation}
%    H_{\rm{crit}} \left( \tau_{\mathrm{p}} \rightarrow \infty \right) \sim \omega_{4,\|}.
%    \label{eq:2}
%\end{equation}
%%%%%%%%%%%%%%%
%This asymptotic expression helps us to understand that when the temperature increases, i.e. the effective macroscopic anisotropy and the energy barrier decrease, the critical field reduces. 

For short pulse duration, $\tau_{\mathrm{p}} \rightarrow 0$, by comparison to the numerical solution of Eq. \eqref{eq:1} it has been suggested that $H_{\mathrm{crit}} \sim \alpha / \tau_{\mathrm{p}}$\cite{roy2016robust}. This inverse scaling with $\tau_{\mathrm{p}}$ strongly limits the use of ultrafast SOT for switching AFM since $H_{\mathrm{crit}}$ would strongly increase for femtosecond time scale stimuli, e.g. for electric fields generated by THz pulses. At the same time, the reduction of $H_{\mathrm{crit}}$ could be achieved by the control of the damping value, however, for a given material the damping value can hardly be modified in a controllable and dynamical way. An alternative to modify $H_{\mathrm{crit}}$ is by regulating the working temperature, and therefore the coefficients entering Eq. \eqref{eq:1}. It is well-known that the exchange and anisotropy constants reduce  by increasing the temperature \cite{atxitia2010multiscale,Evans2020}. However, it is unknown how exactly $H_{\mathrm{crit}}$ depends on the anisotropy and the exchange parameters.

\begin{figure}[!ht]
\includegraphics[width=8.6cm]{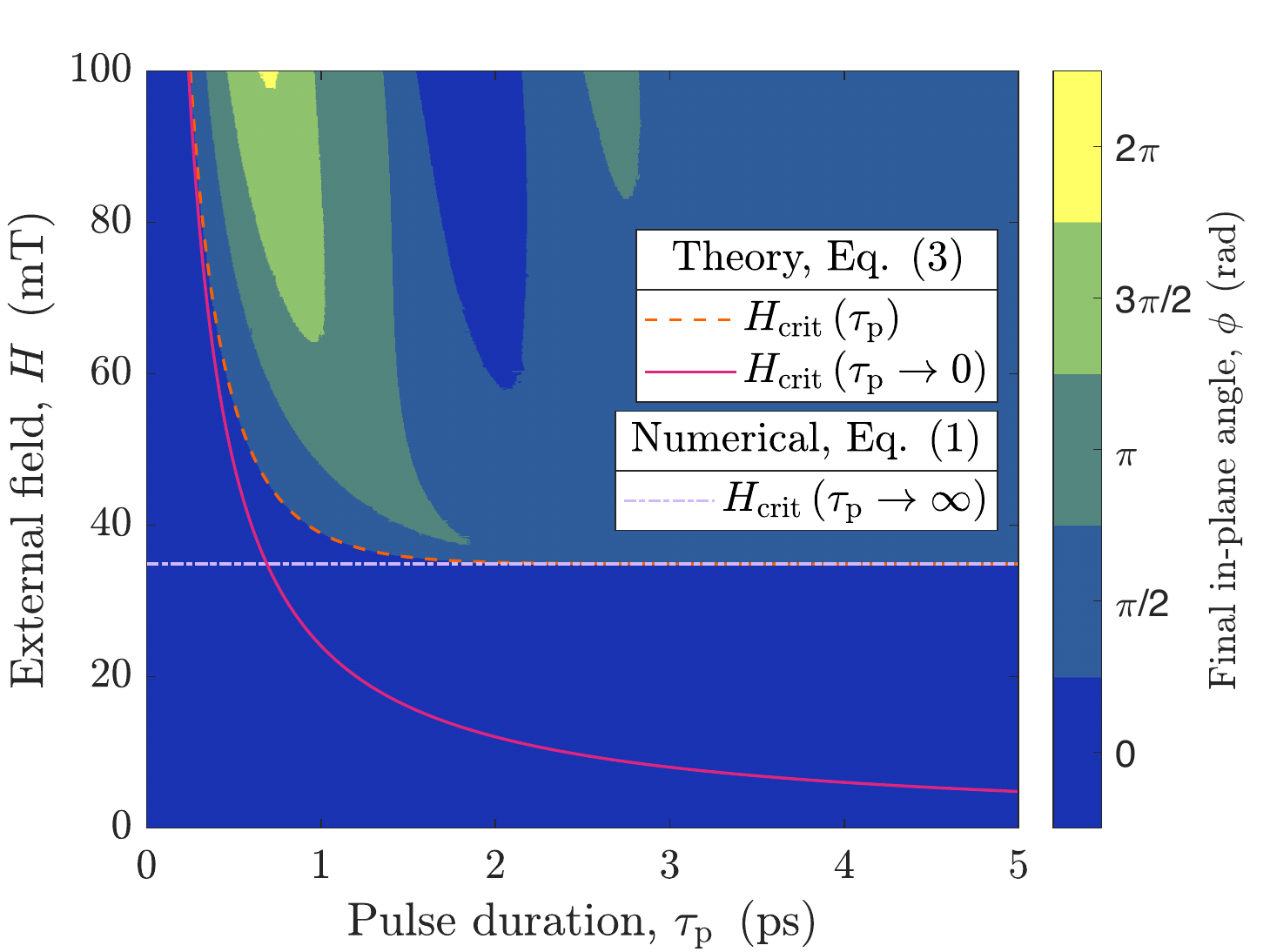}
\caption{Final in-plane angle, $\phi$ as a function of the pulse duration, $\tau_{\rm{p}}$ and the external field $H$. Initial angle, $\phi=0$.}
\label{im:1}
\end{figure}

%The dynamics is essentially given by the balance between SO and damping torque in Eq. \eqref{eq:1}. 
By the numerical solution of Eq. \eqref{eq:1} for values of the damping parameter such that $\alpha \omega_{\mathrm{e}}/\omega_{\mathrm{R}} \ll 1$ ($T=0$ K),  we find that for very short pulse duration, $H_{\rm{crit}}$ scales with the system parameters as follows, 
%%%%%%%%%%%%%%%%
\begin{equation}
    \frac{H_{\rm{crit}} \left( \tau_{\mathrm{p}} \rightarrow 0 \right)}{ H_{\mathrm{crit}} \left( \tau_{\mathrm{p}} \rightarrow \infty \right)}  \sim  \frac{1}{\tau_{\mathrm{p}} \omega_{\mathrm{R}}}.
    \label{eq:3}
\end{equation}
%%%%%%%%%%%%%%%%
It is worth noticing that this expression of the reduced critical field (as normalized by $H_{\mathrm{crit}}\left( \tau_{\mathrm{p}} \rightarrow \infty \right)$) is independent of the damping value.
For instance, when we reduce the anisotropy value from $164.5$ mT used here to $10$ mT used in Ref. \onlinecite{roy2016robust}, we also find that $H_{\mathrm{crit}}$ scales linearly with the damping parameter. 
We also emphasize that for
$\alpha \omega_{\mathrm{e}}/\omega_{\mathrm{R}}\ll 1 $, $H_{\mathrm{crit}} \left( \tau_{\mathrm{p}} \rightarrow \infty \right) \sim \omega_{4,\|}^{\nu}$, with $\nu>1$,  rather than linear ($\nu=1$), which is found when the ratio $\alpha \omega_{\mathrm{e}}/\omega_{\mathrm{R}} \geq 1$. 
We note that $H_{\mathrm{crit}} \left( \tau_{\mathrm{p}} \rightarrow \infty \right)$ slightly depends on the damping value for relatively high anisotropy values used here.  
%The regime, $\alpha \omega_{\mathrm{e}}/\omega_{\mathrm{R}} \sim \alpha\sqrt{\omega_{\mathrm{e}}/\omega_{4,\|}}\ll 1$, can be achieved by either reducing the damping value or  by increasing the value of the tetragonal anisotropy. 
%Notably, we find that as the value of the tetragonal anisotropy increases, $H_{\mathrm{crit}}(\infty)/\omega_{4,\|}$ reduces up to a 25\% from the predicted value. 
%\hl{Although in absolute values higher tetragonal anisotropy means larger critical fields, it also means higher thermal stability. Thus, our findings suggest that antiferromagnetic materials with higher values of the tetragonal anisotropy or heterostructures enhancing such anisotropy should be \texit{looked} after.}

Fig. \ref{im:1} shows a diagram with the switching behaviours under the action of a SOT pulse of duration $\tau_{\mathrm{p}}$ and strength $H$ calculated numerically solving Eq. \eqref{eq:1}. 
Further, by close inspection of the $\tau_{\mathrm{p}}$ dependence of $H_{\rm{crit}}$ in Fig. \ref{im:1}, we find that one can connect the short- and long-duration limits of $H_{\mathrm{crit}}$ by the simplified expression
%%%%%%%%%%%%%%%%%
\begin{equation}
 \frac{H_{\mathrm{crit}}}{ H_{\mathrm{crit}} \left( \tau_{\mathrm{p}} \rightarrow \infty \right)}= \mathrm{coth} \left( \frac{\tau_{\mathrm{p}}}{\tau_{0}}
 \frac{\omega_{\mathrm{R}}}{\omega_{\mathrm{e}}} \right),
\label{eq:4}    
\end{equation}
%%%%%%%%%%%%%%%%
where we find that $\tau_{0}\approx 10$ fs scales as $\tau_0 \sim 1/\omega_{\rm{e}}$. 
Here, the value of $H_{\mathrm{crit}} \left( \tau_{\mathrm{p}} \rightarrow \infty \right)$ is calculated numerically for long enough pulse durations. 
%For $\omega_{4,\|}\rightarrow 0$, it coincides with the prediction in Ref. \onlinecite{roy2016robust}. 
Since $\coth{2} \approx 1$, in Eq. \eqref{eq:4}, for the characteristic pulse duration $\tau_{\mathrm{p}}=2\tau_{0}\omega_{\mathrm{e}} /\omega_{\mathrm{R}}$, one finds that $H_{\mathrm{crit}} \approx H_{\mathrm{crit}} \left( \tau_{\mathrm{p}} \rightarrow \infty \right)$. 
From a direct comparison between Eq. \eqref{eq:4} and the numerical solution of Eq. \eqref{eq:1} for $\alpha \approx 0.001$, we find the value $2\tau_{0}\omega_{\mathrm{e}} /\omega_{\mathrm{R}}=1.4$ ps.
%or $\tau_{\mathrm{p}} > \tau_{\rm{ex}}\omega_{\mathrm{e}} /\omega_{\mathrm{R}}$, $H_{\mathrm{crit}}\rightarrow H_{\mathrm{crit}}(\infty)$, which is constant and related to the tetragonal anisotropy. 
From Fig. \ref{im:1}, one observes that as the pulse duration shortens, $H_{\mathrm{crit}}$ strongly increases. From Eq. \eqref{eq:4}, one realises that the critical field scales approximately inversely with the pulse duration (see Eq. \eqref{eq:1}) for pulse duration shorter than $\tau_{\mathrm{p}}<\tau_{0}\omega_{\mathrm{e}} /\left( \omega_{\mathrm{R}}/2 \right) \approx 350$ fs.  
We note that for a given short-duration pulse, by further increasing the applied field $H \gg H_{\mathrm{crit}}$, the system reaches a region where the inertial effects are so strong that the system could overshoot, i.e. reorient not only $90^{\circ}$ but also $180^{\circ}$ and $270^{\circ}$. A detailed study of this area of the phase diagram is not the scope of this work. Here, our interest centres on the minimum $H_{\mathrm{crit}}$ for a $90^{\circ}$ reorientation, for short-to-intermediate pulse duration and its temperature dependence.  

\begin{figure}[!t]
\includegraphics[width=8.6cm]{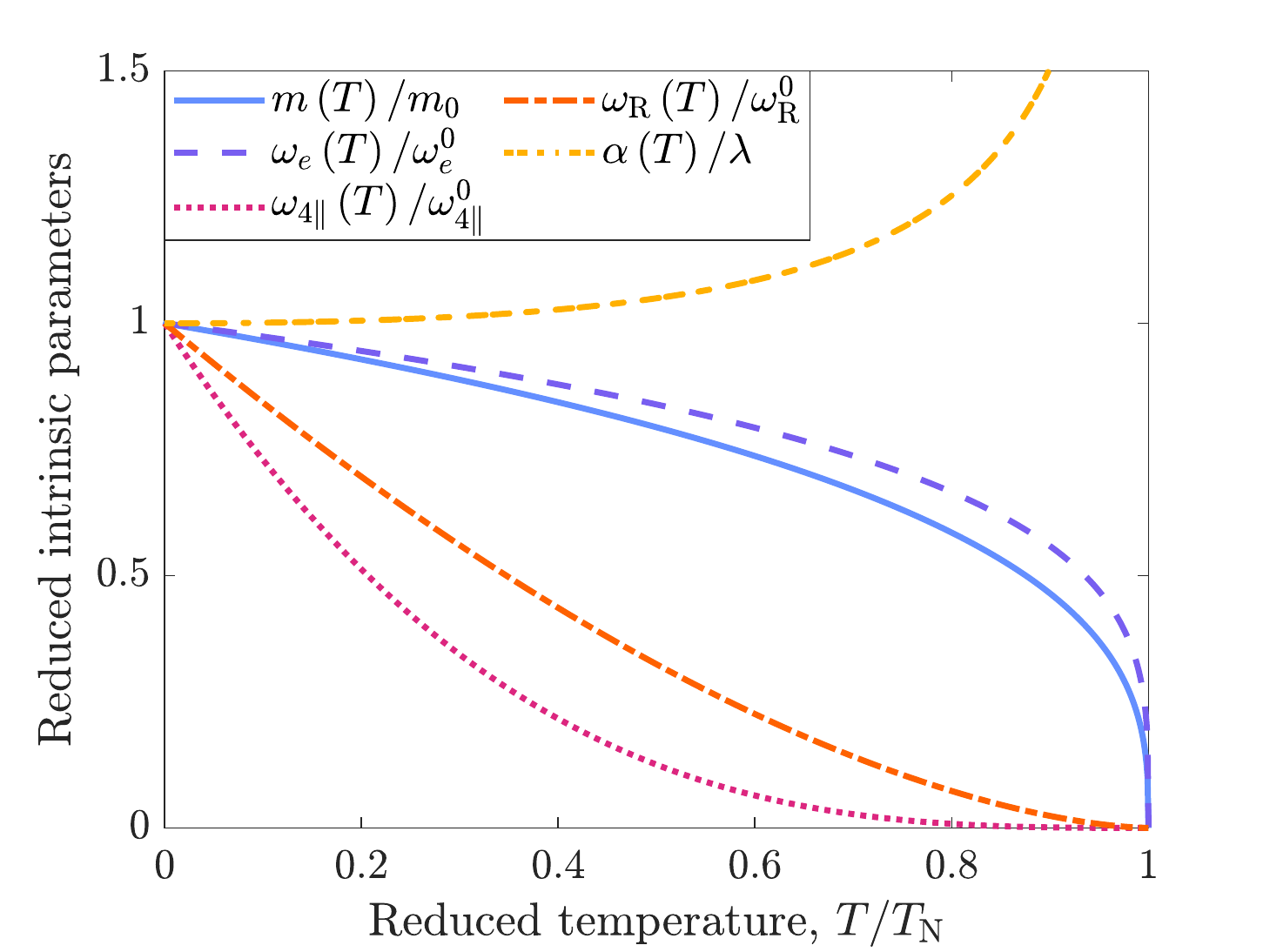}
\caption{Temperature dependence of the system parameters.}
\label{im:2}
\end{figure}

The temperature dependence of the critical field is determined by the temperature dependence of the parameters ($\omega_{\mathrm{e}},\omega_{4,\|},$ and $\alpha$).
Here, we assume that they follow the same scaling laws as the FM systems. The sublattice magnetization modulus, $\abs{\mathbf{m}_i \left( T \right)}=m \left( T \right)$, is given by\cite{evans2015quantitative}
%%%%%%%%%%%%%%%%
\begin{equation}
m \left( T \right)={\left( 1- \frac{T}{T_{\mathrm{N}}} \right)}^{1/3}.
\label{eq:5}
\end{equation}
%%%%%%%%%%%%%%
We also assume that the exchange stiffness, and therefore the exchange frequency, follows the same temperature scaling as in FM \cite{atxitia2010multiscale}, 
%and taking into account that the sublattice atomic moment verifies that $\mu_{\mathrm{at}} \left( T \right) \propto m \left( T \right)$,
%%%%%%%%%%%%%%%%
\begin{equation}
\omega_{\mathrm{e}} \left( T \right)= \omega^0_{\mathrm{e}} \, m^{0.76} \left( T \right).
\label{eq:6}    
\end{equation}
%%%%%%%%%%%%%%%%
%being its value at $T=0$ K specified by $\omega^0_{\mathrm{e}}=\omega_{\mathrm{e}} \left( T =0 \, \mathrm{K} \right)= \left( 1551 \, \mathrm{T} \right) \gamma$. 
Additionally, the temperature dependence of the tetragonal in-plane anisotropy follows the well-known Callen-Callen law \cite{callen1966present}, 
%%%%%%%%%%%%%%%%
\begin{equation}
\omega_{4 \parallel} \left( T \right)= \omega^0_{4 \parallel} \, m^9 \left( T \right).  \label{eq:7} 
\end{equation}
%%%%%%%%%%%%%%
It is precisely this strong temperature dependence of the tetragonal  anisotropy which mainly governs the temperature dependence of the related frequency $\omega_{\mathrm{R}}$, such that $\omega_{\mathrm{R}} \left( T \right)= \omega^0_{\mathrm{R}} \, m^{9.76/2} \left( T \right)$. 
Finally, we assume that the temperature dependence of the macroscopic damping parameter, $\alpha \left( T \right)$, is the same as in FM\cite{Garanin1997,Garanin2004},
%%%%%%%%%%%%%%%
\begin{equation}
\alpha \left( T \right)= \frac{\lambda}{m \left( T \right)} \left( 1- \frac{T}{3 T_{\mathrm{N}}} \right).   
\label{eq:8}    
\end{equation}
%%%%%%%%%%%%%%%
Fig. \ref{im:2} shows the temperature dependence of the reduced parameters -- normalized by their values at zero temperature -- up to N\'eel temperature. One can observe that up to $0.5 \, T_{\mathrm{N}}$, the temperature dependence of the exchange frequency $\omega_{\mathrm{e}}$ and damping $\alpha$ is 
much lower that that of the anisotropy frequency $\omega_{4\|}$ and the related frequency $\omega_{\mathrm{R}}$. Except for the effective damping parameter, all parameters vanish approaching $T_{\mathrm{N}}$. 
For temperature values up to $T=0.5 \, T_{\mathrm{N}}$, the ratio $\alpha \omega_{\mathrm{e}}/\omega_{\mathrm{R}} \ll 1$, thus Eq. \eqref{eq:4} should work. 

We validate our assumptions for the temperature scaling of the system parameters by direct comparison to ASD simulations. The following atomistic spin Hamiltonian describes the same physics as Eq. \eqref{eq:1},
%%%%%%%%%%%%%%%%
\begin{gather}
\mathcal{H}=- \frac{J}{2} \sum_{\left< ij \right>} \mathbf{S}_i \cdot \mathbf{S}_j - \frac{d_{4 \parallel}}{2} \sum_i \left[ {\left( \mathbf{S}_i \cdot \mathbf{\hat{x}} \right)}^4+{\left( \mathbf{S}_i \cdot \mathbf{\hat{y}} \right)}^4 \right] \nonumber \\ -\mu_{\mathrm{at}} \sum_i \mathbf{S}_i \cdot \mathbf{H}_i,
\label{eq:9}
\end{gather}
%%%%%%%%%%%%%%
$\mathbf{S}_i=\boldsymbol{\mu}_i/\mu_{\mathrm{at}}$, being $\boldsymbol{\mu}_i$ the magnetic moment vector at the $i$-{\it th} lattice site. The sum $\left< ij \right>$ runs only over first neighbours of the atomic positions represented through the indices $i$,$j$, whose interaction is represented by the AFM exchange constant $J$ (the same as the macroscopic equation). The second term represents the biaxial fourth-order anisotropy, $d_{4 \parallel}$ (the same as the macroscopic equation), which is defined along the Cartesian $x$- and $y$-{\it th} axes. 
The last contribution of Eq. \eqref{eq:7} represents the Zeeman-like interaction, which in this work is staggered between contiguous antiparallel magnetic sublattices. 
%To investigate the effect of the temperature, $T$, in the system represented by Fig. \ref{im:1} (a), and whose energy is denoted by Eq. \eqref{eq:1}, we will perform ASD simulations in which the thermal effects will be included, in the Langevin dynamics framework, through a white noise term \cite{Atxitia2009}. 

\begin{figure}[!t]
\includegraphics[width=8.6cm]{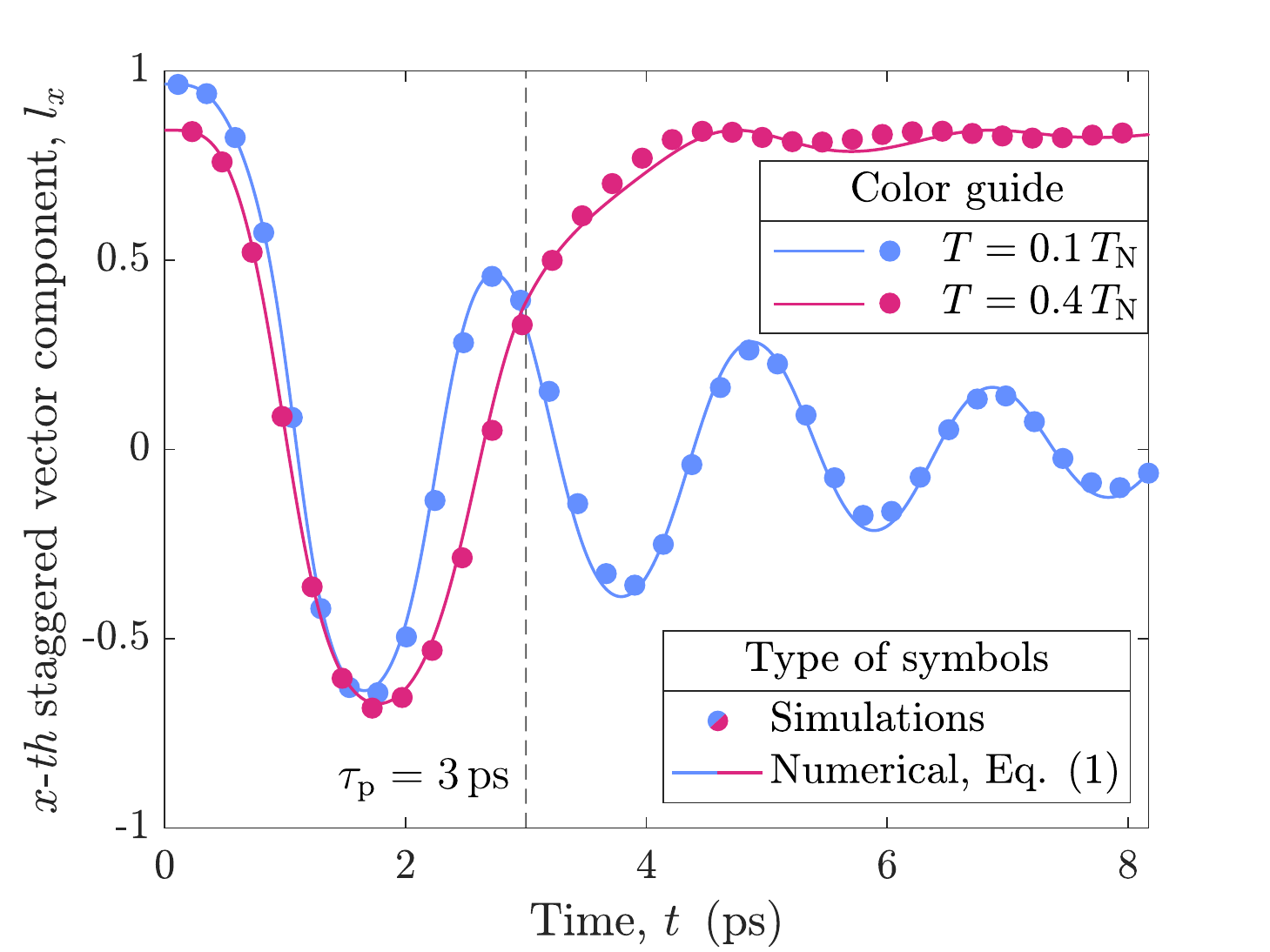}
\caption{Time evolution of the $x$-{\it th} staggered vector component, $l_x$, under the action of a staggered external field of $H=50$ mT for a pulse duration of $\tau_{\mathrm{p}}=3$ ps, for the temperatures $T=0.1 \, T_{\mathrm{N}}$ and $0.4 \, T_{\mathrm{N}}$, being $T_{\mathrm{N}}$ the N\'eel temperature. Symbols correspond to ASD simulations and lines to the numerical solution of Eq. \eqref{eq:1} with temperature-dependent parameters.}
\label{im:3}
\end{figure}

The spin dynamics at finite temperatures is described by the atomistic stochastic Landau-Lifshitz-Gilbert (LLG) equation,
%%%%%%%%%%%%%%%%%
\begin{equation}
{\dot{\mathbf{S}}}_i=-\frac{\gamma}{\left( 1+ \lambda^2 \right)} \, \mathbf{S}_i \times \left[ \mathbf{H}^{\mathrm{eff}}_i+\lambda \left( \mathbf{S}_i \times \mathbf{H}^{\mathrm{eff}}_i \right) \right],    
\label{eq:10}    
\end{equation}
%%%%%%%%%%%%%%
 here, $\gamma$ is the gyromagnetic ratio and $\lambda$ the phenomenological atomistic damping constant, and where $\mathbf{H}^{\mathrm{eff}}_i = - \left( 1/\mu_{\mathrm{at}} \right) \left( \partial \mathcal{H} / \delta \mathbf{S}_i \right)+\boldsymbol{\xi}_i$ symbolizes the effective field at the $i$-{\it th} lattice position, where $\boldsymbol{\xi}_i$ represents the associated stochastic field\cite{Atxitia2009}. We conduct the following numerical experiment using both Eq. \eqref{eq:1} and ASD simulations. As initial state, we select that the sublattice magnetization lies along the $\pm x$-{\it th} easy-axis in the sublattices A and B, respectively. We switch on the thermostat in the ASD simulation and let the system relax to its thermodynamic equilibrium  $\mathbf{l}=(l_{\mathrm{eq}},0,0) \rightarrow \phi = 0$, where $l_{\mathrm{eq}}$ is the equilibrium value at a given temperature  (see Fig. \ref{im:2}). At $t=0$, we apply an external field along the $y$-{\it th} spatial direction, with alternating sign between spin sublattices, such that $\mathbf{H}_i = \pm H \, \mathbf {\hat{y}}$, where $H$ is the magnitude of the field.
 %, as denoted in Fig. \ref{im:1} (a). 
 %We explore the dynamics of the N\'eel order parameter, $\mathbf{l} = \left( \mathbf{m}_{\mathrm{A}} - \mathbf{m}_{\mathrm{B}} \right)/2$, where $\mathbf{m}_i$ denotes the macroscopic reduced magnetization vector at the $i$-{\it th} magnetic sublattice. 
 We compare the dynamics of $l_x$ for $\lambda=0.001$, a pulse duration $\tau_{\mathrm{p}} = 3 $ ps, a field $ H = 50 $ mT, for a range of temperatures. The numerical calculations using Eq. \eqref{eq:1} include the temperature scaling of the parameters described above. In Fig. \ref{im:3}  we illustrate for two temperatures ($T=0.1 \, T_{\mathrm{N}}$ and $0.4 \, T_{\mathrm{N}}$), the good agreement  between ASD simulations and theory. 
% We compare the dynamics of $l_x$ calculated by semi-analytical model (Eq. \eqref{eq:1}) and ASD simulations for $\lambda=0.001$, a pulse duration $\tau_{\mathrm{p}} = 3 $ ps, a field $ H = 50 $ mT, for two temperatures, $T=0.1 \, T_{\mathrm{N}}$ and $0.4 \, T_{\mathrm{N}}$. 
For a temperature of $T=0.1 \, T_{\mathrm{N}}$ magnetic order switches $\pi/2$ ($l_x=0$). For a higher temperature, $T=0.4 \, T_{\mathrm{N}}$,
$l_x$ follows a similar dynamics while the stimulus is on ($t<\tau_{\mathrm{p}}$), however, once it is turned off, the magnetic order relaxes back to the initial state $\phi=0$ (see Fig. \ref{im:3}).

\begin{figure}[!t]
\includegraphics[width=8.6cm]{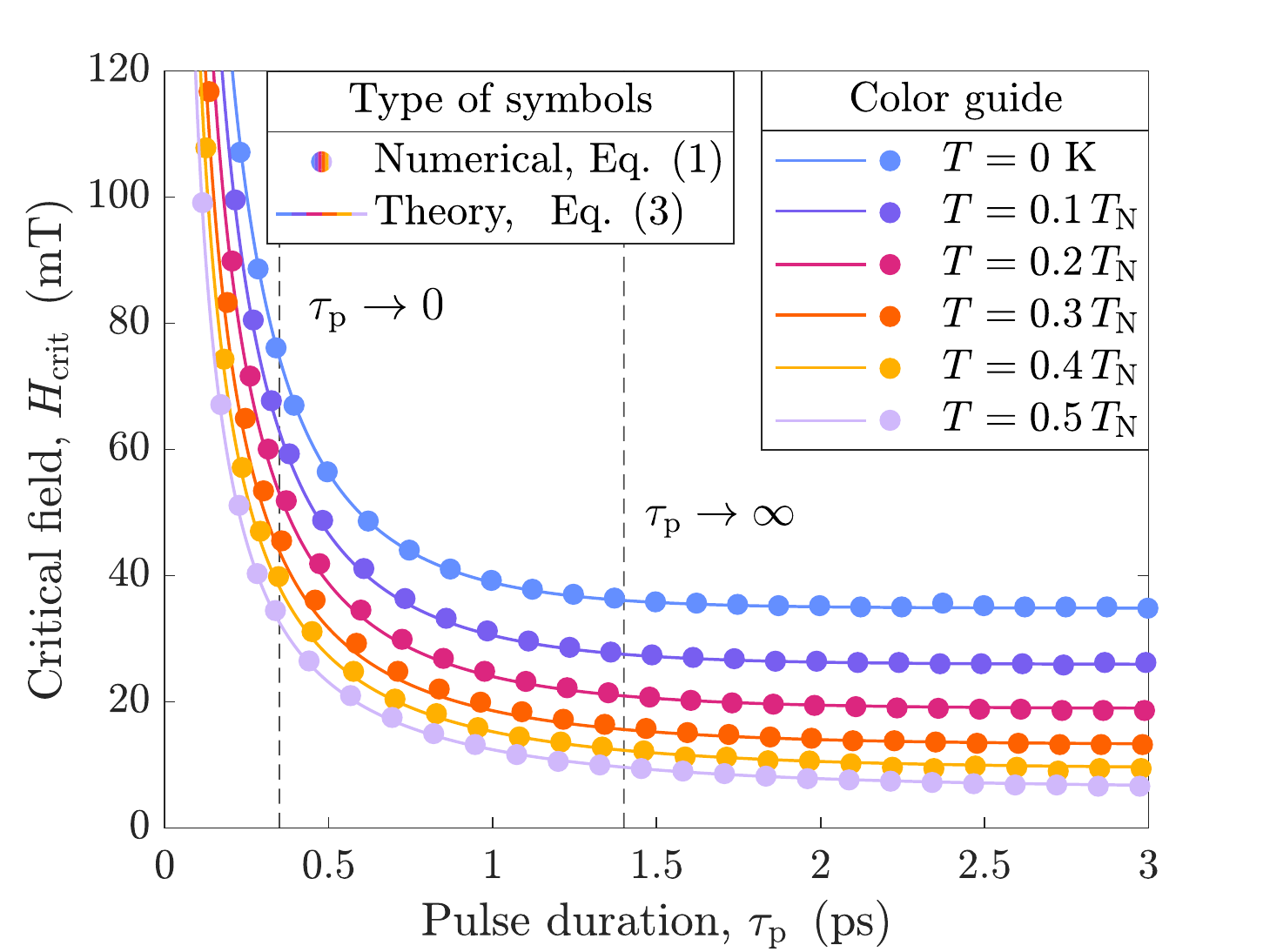}
\caption{Symbols correspond to the critical field calculated from the numerical solution of Eq. \eqref{eq:1}. Lines correspond to the analytical function in Eq. \eqref{eq:4}.}
\label{im:4}
\end{figure}

We verify that the extension of Eq. \eqref{eq:4} to finite temperatures is valid by estimating numerically $H_{\rm{crit}}$ for a range of temperature (from $T=0$ K to $0.5 \, T_\mathrm{N}$) and pulse duration ($\tau_{\mathrm{p}}=0.1-20$ ps). 
For each temperature and pulse duration we vary the value of $H$ until we find the minimum value, $H_{\rm{crit}}$, for which the  final state of the system has switched by $90^{\circ}$. 
Fig. \ref{im:4} shows $H_{\rm{crit}}$ for a range of temperatures calculated by the numerical solution of Eq. \eqref{eq:1} (symbols) and the analytical Eq. \eqref{eq:4} (solid lines), where $\omega_{\mathrm{e}} \sim m^{0.76}$ and $\omega_{\mathrm{R}} \sim m^{9.76/2}$. 
For the limit, $\tau_{\mathrm{p}} \rightarrow \infty$, we observe that $H_{\rm{crit}}$ strongly reduces.
For the system parameters used here, the critical field reduces from $34.8$ mT ($T=0$ K) to $6.4$ mT ($T=0.5 \, T_{\rm{N}}$), and the power dissipated as Joule heating by a factor larger than $2$. 
Although this route (using long duration pulses) is the most efficient for reducing the critical field, it limits its applicability to stimuli longer a characteristic time scale $\sim  2\tau_{0} \omega_{\mathrm{e}}/\omega_{\mathrm{R}}$. 
For very short pulses, the reduction is less dramatic, but still,  for e.g. $\tau_{\mathrm{p}}=140$ fs, the effective field can be reduced by a factor 2 (see Fig. \ref{im:4}). This means that the power necessary to reach the critical current and consequently the critical switching field, would also reduce, which could be critical, between sample damage or not. 
Still, it remains to solve the practical issue of achieving electric currents of only a few hundred of femtoseconds \cite{jhuria2020spin}. One could also explore the effect of temperature when the electric field of THz radiation is used as stimulus\cite{olejnik2018terahertz}. Our results could also be used to devise strategies to reduce the critical field such as heat-assisted magnetic switching, where a laser pulse transiently heats up the system, reducing the critical field.  To explore such scenario, one would need to go beyond the LLG approach used here, which conserves the magnetic order length, and use for instance the Landau-Lifshitz-Bloch model\cite{atxitia2016fundamentals,hirst2022multiscale}.

%The authors have no conflicts to disclose.

U.A. and R.R.-E. acknowledge support from the Deutsche Forschungsgemeinschaft through SFB/TT 227 ``Ultrafast spin dynamics", Project A08. The work of R.M.O. was partially supported by the STSM grants from the COST Action CA17123 ``Ultrafast opto-magneto-electronics for non-dissipative information technology".

%The data that support the findings of this study are available from the corresponding authors upon reasonable request.

\bibliographystyle{apsrev4-1}
\bibliography{Bibliography}

\begin{thebibliography}{28}%
\makeatletter
\providecommand \@ifxundefined [1]{%
 \@ifx{#1\undefined}
}%
\providecommand \@ifnum [1]{%
 \ifnum #1\expandafter \@firstoftwo
 \else \expandafter \@secondoftwo
 \fi
}%
\providecommand \@ifx [1]{%
 \ifx #1\expandafter \@firstoftwo
 \else \expandafter \@secondoftwo
 \fi
}%
\providecommand \natexlab [1]{#1}%
\providecommand \enquote  [1]{``#1''}%
\providecommand \bibnamefont  [1]{#1}%
\providecommand \bibfnamefont [1]{#1}%
\providecommand \citenamefont [1]{#1}%
\providecommand \href@noop [0]{\@secondoftwo}%
\providecommand \href [0]{\begingroup \@sanitize@url \@href}%
\providecommand \@href[1]{\@@startlink{#1}\@@href}%
\providecommand \@@href[1]{\endgroup#1\@@endlink}%
\providecommand \@sanitize@url [0]{\catcode `\\12\catcode `\$12\catcode
  `\&12\catcode `\#12\catcode `\^12\catcode `\_12\catcode `\%12\relax}%
\providecommand \@@startlink[1]{}%
\providecommand \@@endlink[0]{}%
\providecommand \url  [0]{\begingroup\@sanitize@url \@url }%
\providecommand \@url [1]{\endgroup\@href {#1}{\urlprefix }}%
\providecommand \urlprefix  [0]{URL }%
\providecommand \Eprint [0]{\href }%
\providecommand \doibase [0]{http://dx.doi.org/}%
\providecommand \selectlanguage [0]{\@gobble}%
\providecommand \bibinfo  [0]{\@secondoftwo}%
\providecommand \bibfield  [0]{\@secondoftwo}%
\providecommand \translation [1]{[#1]}%
\providecommand \BibitemOpen [0]{}%
\providecommand \bibitemStop [0]{}%
\providecommand \bibitemNoStop [0]{.\EOS\space}%
\providecommand \EOS [0]{\spacefactor3000\relax}%
\providecommand \BibitemShut  [1]{\csname bibitem#1\endcsname}%
\let\auto@bib@innerbib\@empty
%</preamble>
\bibitem [{\citenamefont {Wadley}\ \emph {et~al.}(2016)\citenamefont {Wadley},
  \citenamefont {Howells}, \citenamefont {{\v{Z}}elezn{\`y}}, \citenamefont
  {Andrews}, \citenamefont {Hills}, \citenamefont {Campion}, \citenamefont
  {Nov{\'a}k}, \citenamefont {Olejn{\'\i}k}, \citenamefont {Maccherozzi},
  \citenamefont {Dhesi} \emph {et~al.}}]{wadley2016electrical}%
  \BibitemOpen
  \bibfield  {author} {\bibinfo {author} {\bibfnamefont {P.}~\bibnamefont
  {Wadley}}, \bibinfo {author} {\bibfnamefont {B.}~\bibnamefont {Howells}},
  \bibinfo {author} {\bibfnamefont {J.}~\bibnamefont {{\v{Z}}elezn{\`y}}},
  \bibinfo {author} {\bibfnamefont {C.}~\bibnamefont {Andrews}}, \bibinfo
  {author} {\bibfnamefont {V.}~\bibnamefont {Hills}}, \bibinfo {author}
  {\bibfnamefont {R.~P.}\ \bibnamefont {Campion}}, \bibinfo {author}
  {\bibfnamefont {V.}~\bibnamefont {Nov{\'a}k}}, \bibinfo {author}
  {\bibfnamefont {K.}~\bibnamefont {Olejn{\'\i}k}}, \bibinfo {author}
  {\bibfnamefont {F.}~\bibnamefont {Maccherozzi}}, \bibinfo {author}
  {\bibfnamefont {S.}~\bibnamefont {Dhesi}},  \emph {et~al.},\ }\href
  {https://www.science.org/doi/full/10.1126/science.aab1031?casa_token=5D0OdFiMb34AAAAA%3AGnDnvtm4DY4WHol3TJ9vA10qTNN51fxOQAzIZ1GpeZZKWQ_0YKnXMdmOkZcPOppRqKriN_HVMnbr}
  {\bibfield  {journal} {\bibinfo  {journal} {Science}\ }\textbf {\bibinfo
  {volume} {351}},\ \bibinfo {pages} {587} (\bibinfo {year}
  {2016})}\BibitemShut {NoStop}%
\bibitem [{\citenamefont {Janda}\ \emph {et~al.}(2020)\citenamefont {Janda},
  \citenamefont {Godinho}, \citenamefont {Ostatnicky}, \citenamefont
  {Pfitzner}, \citenamefont {Ulrich}, \citenamefont {Hoehl}, \citenamefont
  {Reimers}, \citenamefont {{\v{S}}ob{\'a}{\v{n}}}, \citenamefont {Metzger},
  \citenamefont {Reichlova} \emph {et~al.}}]{janda2020magneto}%
  \BibitemOpen
  \bibfield  {author} {\bibinfo {author} {\bibfnamefont {T.}~\bibnamefont
  {Janda}}, \bibinfo {author} {\bibfnamefont {J.}~\bibnamefont {Godinho}},
  \bibinfo {author} {\bibfnamefont {T.}~\bibnamefont {Ostatnicky}}, \bibinfo
  {author} {\bibfnamefont {E.}~\bibnamefont {Pfitzner}}, \bibinfo {author}
  {\bibfnamefont {G.}~\bibnamefont {Ulrich}}, \bibinfo {author} {\bibfnamefont
  {A.}~\bibnamefont {Hoehl}}, \bibinfo {author} {\bibfnamefont
  {S.}~\bibnamefont {Reimers}}, \bibinfo {author} {\bibfnamefont
  {Z.}~\bibnamefont {{\v{S}}ob{\'a}{\v{n}}}}, \bibinfo {author} {\bibfnamefont
  {T.}~\bibnamefont {Metzger}}, \bibinfo {author} {\bibfnamefont
  {H.}~\bibnamefont {Reichlova}},  \emph {et~al.},\ }\href
  {https://journals.aps.org/prmaterials/abstract/10.1103/PhysRevMaterials.4.094413}
  {\bibfield  {journal} {\bibinfo  {journal} {Phys. Rev. Mater.}\ }\textbf
  {\bibinfo {volume} {4}},\ \bibinfo {pages} {094413} (\bibinfo {year}
  {2020})}\BibitemShut {NoStop}%
\bibitem [{\citenamefont {Bodnar}\ \emph {et~al.}(2018)\citenamefont {Bodnar},
  \citenamefont {{\v{S}}mejkal}, \citenamefont {Turek}, \citenamefont
  {Jungwirth}, \citenamefont {Gomonay}, \citenamefont {Sinova}, \citenamefont
  {Sapozhnik}, \citenamefont {Elmers}, \citenamefont {Kl{\"a}ui},\ and\
  \citenamefont {Jourdan}}]{bodnar2018writing}%
  \BibitemOpen
  \bibfield  {author} {\bibinfo {author} {\bibfnamefont {S.~Y.}\ \bibnamefont
  {Bodnar}}, \bibinfo {author} {\bibfnamefont {L.}~\bibnamefont
  {{\v{S}}mejkal}}, \bibinfo {author} {\bibfnamefont {I.}~\bibnamefont
  {Turek}}, \bibinfo {author} {\bibfnamefont {T.}~\bibnamefont {Jungwirth}},
  \bibinfo {author} {\bibfnamefont {O.}~\bibnamefont {Gomonay}}, \bibinfo
  {author} {\bibfnamefont {J.}~\bibnamefont {Sinova}}, \bibinfo {author}
  {\bibfnamefont {A.~A.}\ \bibnamefont {Sapozhnik}}, \bibinfo {author}
  {\bibfnamefont {H.-J.}\ \bibnamefont {Elmers}}, \bibinfo {author}
  {\bibfnamefont {M.}~\bibnamefont {Kl{\"a}ui}}, \ and\ \bibinfo {author}
  {\bibfnamefont {M.}~\bibnamefont {Jourdan}},\ }\href
  {https://www.nature.com/articles/s41467-017-02780-x} {\bibfield  {journal}
  {\bibinfo  {journal} {Nat. Commun.}\ }\textbf {\bibinfo {volume} {9}},\
  \bibinfo {pages} {1} (\bibinfo {year} {2018})}\BibitemShut {NoStop}%
\bibitem [{\citenamefont {Meinert}\ \emph {et~al.}(2018)\citenamefont
  {Meinert}, \citenamefont {Graulich},\ and\ \citenamefont
  {Matalla-Wagner}}]{meinert2018electrical}%
  \BibitemOpen
  \bibfield  {author} {\bibinfo {author} {\bibfnamefont {M.}~\bibnamefont
  {Meinert}}, \bibinfo {author} {\bibfnamefont {D.}~\bibnamefont {Graulich}}, \
  and\ \bibinfo {author} {\bibfnamefont {T.}~\bibnamefont {Matalla-Wagner}},\
  }\href
  {https://journals.aps.org/prapplied/abstract/10.1103/PhysRevApplied.9.064040}
  {\bibfield  {journal} {\bibinfo  {journal} {Phys. Rev. Appl.}\ }\textbf
  {\bibinfo {volume} {9}},\ \bibinfo {pages} {064040} (\bibinfo {year}
  {2018})}\BibitemShut {NoStop}%
\bibitem [{\citenamefont {Godinho}\ \emph {et~al.}(2018)\citenamefont
  {Godinho}, \citenamefont {Reichlov{\'a}}, \citenamefont {Kriegner},
  \citenamefont {Nov{\'a}k}, \citenamefont {Olejn{\'\i}k}, \citenamefont
  {Ka{\v{s}}par}, \citenamefont {{\v{S}}ob{\'a}{\v{n}}}, \citenamefont
  {Wadley}, \citenamefont {Campion}, \citenamefont {Otxoa} \emph
  {et~al.}}]{godinho2018electrically}%
  \BibitemOpen
  \bibfield  {author} {\bibinfo {author} {\bibfnamefont {J.}~\bibnamefont
  {Godinho}}, \bibinfo {author} {\bibfnamefont {H.}~\bibnamefont
  {Reichlov{\'a}}}, \bibinfo {author} {\bibfnamefont {D.}~\bibnamefont
  {Kriegner}}, \bibinfo {author} {\bibfnamefont {V.}~\bibnamefont {Nov{\'a}k}},
  \bibinfo {author} {\bibfnamefont {K.}~\bibnamefont {Olejn{\'\i}k}}, \bibinfo
  {author} {\bibfnamefont {Z.}~\bibnamefont {Ka{\v{s}}par}}, \bibinfo {author}
  {\bibfnamefont {Z.}~\bibnamefont {{\v{S}}ob{\'a}{\v{n}}}}, \bibinfo {author}
  {\bibfnamefont {P.}~\bibnamefont {Wadley}}, \bibinfo {author} {\bibfnamefont
  {R.~P.}\ \bibnamefont {Campion}}, \bibinfo {author} {\bibfnamefont {R.~M.}\
  \bibnamefont {Otxoa}},  \emph {et~al.},\ }\href
  {https://www.nature.com/articles/s41467-018-07092-2} {\bibfield  {journal}
  {\bibinfo  {journal} {Nat. Commun.}\ }\textbf {\bibinfo {volume} {9}},\
  \bibinfo {pages} {1} (\bibinfo {year} {2018})}\BibitemShut {NoStop}%
\bibitem [{\citenamefont {Zhou}\ \emph {et~al.}(2018)\citenamefont {Zhou},
  \citenamefont {Zhang}, \citenamefont {Li}, \citenamefont {Chen},
  \citenamefont {Shi}, \citenamefont {Tan}, \citenamefont {Gu}, \citenamefont
  {Saleem}, \citenamefont {Wu}, \citenamefont {Pan},\ and\ \citenamefont
  {Song}}]{Zhou2018}%
  \BibitemOpen
  \bibfield  {author} {\bibinfo {author} {\bibfnamefont {X.~F.}\ \bibnamefont
  {Zhou}}, \bibinfo {author} {\bibfnamefont {J.}~\bibnamefont {Zhang}},
  \bibinfo {author} {\bibfnamefont {F.}~\bibnamefont {Li}}, \bibinfo {author}
  {\bibfnamefont {X.~Z.}\ \bibnamefont {Chen}}, \bibinfo {author}
  {\bibfnamefont {G.~Y.}\ \bibnamefont {Shi}}, \bibinfo {author} {\bibfnamefont
  {Y.~Z.}\ \bibnamefont {Tan}}, \bibinfo {author} {\bibfnamefont {Y.~D.}\
  \bibnamefont {Gu}}, \bibinfo {author} {\bibfnamefont {M.~S.}\ \bibnamefont
  {Saleem}}, \bibinfo {author} {\bibfnamefont {H.~Q.}\ \bibnamefont {Wu}},
  \bibinfo {author} {\bibfnamefont {F.}~\bibnamefont {Pan}}, \ and\ \bibinfo
  {author} {\bibfnamefont {C.}~\bibnamefont {Song}},\ }\href {\doibase
  10.1103/PhysRevApplied.9.054028} {\bibfield  {journal} {\bibinfo  {journal}
  {Phys. Rev. Appl.}\ }\textbf {\bibinfo {volume} {9}},\ \bibinfo {pages}
  {054028} (\bibinfo {year} {2018})}\BibitemShut {NoStop}%
\bibitem [{\citenamefont {{\v{Z}}elezn{\`y}}\ \emph {et~al.}(2014)\citenamefont
  {{\v{Z}}elezn{\`y}}, \citenamefont {Gao}, \citenamefont {V{\`y}born{\`y}},
  \citenamefont {Zemen}, \citenamefont {Ma{\v{s}}ek}, \citenamefont {Manchon},
  \citenamefont {Wunderlich}, \citenamefont {Sinova},\ and\ \citenamefont
  {Jungwirth}}]{vzelezny2014relativistic}%
  \BibitemOpen
  \bibfield  {author} {\bibinfo {author} {\bibfnamefont {J.}~\bibnamefont
  {{\v{Z}}elezn{\`y}}}, \bibinfo {author} {\bibfnamefont {H.}~\bibnamefont
  {Gao}}, \bibinfo {author} {\bibfnamefont {K.}~\bibnamefont
  {V{\`y}born{\`y}}}, \bibinfo {author} {\bibfnamefont {J.}~\bibnamefont
  {Zemen}}, \bibinfo {author} {\bibfnamefont {J.}~\bibnamefont {Ma{\v{s}}ek}},
  \bibinfo {author} {\bibfnamefont {A.}~\bibnamefont {Manchon}}, \bibinfo
  {author} {\bibfnamefont {J.}~\bibnamefont {Wunderlich}}, \bibinfo {author}
  {\bibfnamefont {J.}~\bibnamefont {Sinova}}, \ and\ \bibinfo {author}
  {\bibfnamefont {T.}~\bibnamefont {Jungwirth}},\ }\href
  {https://journals.aps.org/prl/abstract/10.1103/PhysRevLett.113.157201}
  {\bibfield  {journal} {\bibinfo  {journal} {Phys. Rev. Lett.}\ }\textbf
  {\bibinfo {volume} {113}},\ \bibinfo {pages} {157201} (\bibinfo {year}
  {2014})}\BibitemShut {NoStop}%
\bibitem [{\citenamefont {Chen}\ \emph {et~al.}(2018)\citenamefont {Chen},
  \citenamefont {Zarzuela}, \citenamefont {Zhang}, \citenamefont {Song},
  \citenamefont {Zhou}, \citenamefont {Shi}, \citenamefont {Li}, \citenamefont
  {Zhou}, \citenamefont {Jiang}, \citenamefont {Pan} \emph
  {et~al.}}]{Chen2018}%
  \BibitemOpen
  \bibfield  {author} {\bibinfo {author} {\bibfnamefont {X.~Z.}\ \bibnamefont
  {Chen}}, \bibinfo {author} {\bibfnamefont {R.}~\bibnamefont {Zarzuela}},
  \bibinfo {author} {\bibfnamefont {J.}~\bibnamefont {Zhang}}, \bibinfo
  {author} {\bibfnamefont {C.}~\bibnamefont {Song}}, \bibinfo {author}
  {\bibfnamefont {X.~F.}\ \bibnamefont {Zhou}}, \bibinfo {author}
  {\bibfnamefont {G.~Y.}\ \bibnamefont {Shi}}, \bibinfo {author} {\bibfnamefont
  {F.}~\bibnamefont {Li}}, \bibinfo {author} {\bibfnamefont {H.~A.}\
  \bibnamefont {Zhou}}, \bibinfo {author} {\bibfnamefont {W.~J.}\ \bibnamefont
  {Jiang}}, \bibinfo {author} {\bibfnamefont {F.}~\bibnamefont {Pan}},  \emph
  {et~al.},\ }\href
  {https://journals.aps.org/prl/abstract/10.1103/PhysRevLett.120.207204}
  {\bibfield  {journal} {\bibinfo  {journal} {Phys. Rev. Lett.}\ }\textbf
  {\bibinfo {volume} {120}},\ \bibinfo {pages} {207204} (\bibinfo {year}
  {2018})}\BibitemShut {NoStop}%
\bibitem [{\citenamefont {Chirac}\ \emph {et~al.}(2020)\citenamefont {Chirac},
  \citenamefont {Chauleau}, \citenamefont {Thibaudeau}, \citenamefont
  {Gomonay},\ and\ \citenamefont {Viret}}]{Chirac2020}%
  \BibitemOpen
  \bibfield  {author} {\bibinfo {author} {\bibfnamefont {T.}~\bibnamefont
  {Chirac}}, \bibinfo {author} {\bibfnamefont {J.-Y.}\ \bibnamefont
  {Chauleau}}, \bibinfo {author} {\bibfnamefont {P.}~\bibnamefont
  {Thibaudeau}}, \bibinfo {author} {\bibfnamefont {O.}~\bibnamefont {Gomonay}},
  \ and\ \bibinfo {author} {\bibfnamefont {M.}~\bibnamefont {Viret}},\ }\href
  {https://journals.aps.org/prb/abstract/10.1103/PhysRevB.102.134415}
  {\bibfield  {journal} {\bibinfo  {journal} {Phys. Rev. B}\ }\textbf {\bibinfo
  {volume} {102}},\ \bibinfo {pages} {134415} (\bibinfo {year}
  {2020})}\BibitemShut {NoStop}%
\bibitem [{\citenamefont {Olejn{\'\i}k}\ \emph {et~al.}(2018)\citenamefont
  {Olejn{\'\i}k}, \citenamefont {Seifert}, \citenamefont {Ka{\v{s}}par},
  \citenamefont {Nov{\'a}k}, \citenamefont {Wadley}, \citenamefont {Campion},
  \citenamefont {Baumgartner}, \citenamefont {Gambardella}, \citenamefont
  {N{\v{e}}mec}, \citenamefont {Wunderlich} \emph
  {et~al.}}]{olejnik2018terahertz}%
  \BibitemOpen
  \bibfield  {author} {\bibinfo {author} {\bibfnamefont {K.}~\bibnamefont
  {Olejn{\'\i}k}}, \bibinfo {author} {\bibfnamefont {T.}~\bibnamefont
  {Seifert}}, \bibinfo {author} {\bibfnamefont {Z.}~\bibnamefont
  {Ka{\v{s}}par}}, \bibinfo {author} {\bibfnamefont {V.}~\bibnamefont
  {Nov{\'a}k}}, \bibinfo {author} {\bibfnamefont {P.}~\bibnamefont {Wadley}},
  \bibinfo {author} {\bibfnamefont {R.~P.}\ \bibnamefont {Campion}}, \bibinfo
  {author} {\bibfnamefont {M.}~\bibnamefont {Baumgartner}}, \bibinfo {author}
  {\bibfnamefont {P.}~\bibnamefont {Gambardella}}, \bibinfo {author}
  {\bibfnamefont {P.}~\bibnamefont {N{\v{e}}mec}}, \bibinfo {author}
  {\bibfnamefont {J.}~\bibnamefont {Wunderlich}},  \emph {et~al.},\ }\href
  {https://www.science.org/doi/10.1126/sciadv.aar3566} {\bibfield  {journal}
  {\bibinfo  {journal} {Sci. Adv.}\ }\textbf {\bibinfo {volume} {4}},\ \bibinfo
  {pages} {eaar3566} (\bibinfo {year} {2018})}\BibitemShut {NoStop}%
\bibitem [{\citenamefont {Ka{\v{s}}par}\ \emph {et~al.}(2021)\citenamefont
  {Ka{\v{s}}par}, \citenamefont {Sur{\`y}nek}, \citenamefont {Zub{\'a}{\v{c}}},
  \citenamefont {Krizek}, \citenamefont {Nov{\'a}k}, \citenamefont {Campion},
  \citenamefont {W{\"o}rnle}, \citenamefont {Gambardella}, \citenamefont
  {Marti}, \citenamefont {N{\v{e}}mec} \emph {et~al.}}]{Kaspar2020}%
  \BibitemOpen
  \bibfield  {author} {\bibinfo {author} {\bibfnamefont {Z.}~\bibnamefont
  {Ka{\v{s}}par}}, \bibinfo {author} {\bibfnamefont {M.}~\bibnamefont
  {Sur{\`y}nek}}, \bibinfo {author} {\bibfnamefont {J.}~\bibnamefont
  {Zub{\'a}{\v{c}}}}, \bibinfo {author} {\bibfnamefont {F.}~\bibnamefont
  {Krizek}}, \bibinfo {author} {\bibfnamefont {V.}~\bibnamefont {Nov{\'a}k}},
  \bibinfo {author} {\bibfnamefont {R.~P.}\ \bibnamefont {Campion}}, \bibinfo
  {author} {\bibfnamefont {M.~S.}\ \bibnamefont {W{\"o}rnle}}, \bibinfo
  {author} {\bibfnamefont {P.}~\bibnamefont {Gambardella}}, \bibinfo {author}
  {\bibfnamefont {X.}~\bibnamefont {Marti}}, \bibinfo {author} {\bibfnamefont
  {P.}~\bibnamefont {N{\v{e}}mec}},  \emph {et~al.},\ }\href
  {https://www.nature.com/articles/s41928-020-00506-4} {\bibfield  {journal}
  {\bibinfo  {journal} {Nat. Electron.}\ }\textbf {\bibinfo {volume} {4}},\
  \bibinfo {pages} {30} (\bibinfo {year} {2021})}\BibitemShut {NoStop}%
\bibitem [{\citenamefont {Roy}\ \emph {et~al.}(2016)\citenamefont {Roy},
  \citenamefont {Otxoa},\ and\ \citenamefont {Wunderlich}}]{roy2016robust}%
  \BibitemOpen
  \bibfield  {author} {\bibinfo {author} {\bibfnamefont {P.~E.}\ \bibnamefont
  {Roy}}, \bibinfo {author} {\bibfnamefont {R.~M.}\ \bibnamefont {Otxoa}}, \
  and\ \bibinfo {author} {\bibfnamefont {J.}~\bibnamefont {Wunderlich}},\
  }\href {https://journals.aps.org/prb/abstract/10.1103/PhysRevB.94.014439}
  {\bibfield  {journal} {\bibinfo  {journal} {Phys. Rev. B}\ }\textbf {\bibinfo
  {volume} {94}},\ \bibinfo {pages} {014439} (\bibinfo {year}
  {2016})}\BibitemShut {NoStop}%
\bibitem [{\citenamefont {Kryder}\ \emph {et~al.}(2008)\citenamefont {Kryder},
  \citenamefont {Gage}, \citenamefont {McDaniel}, \citenamefont {Challener},
  \citenamefont {Rottmayer}, \citenamefont {Ju}, \citenamefont {Hsia},\ and\
  \citenamefont {Erden}}]{Kryder2008}%
  \BibitemOpen
  \bibfield  {author} {\bibinfo {author} {\bibfnamefont {M.~H.}\ \bibnamefont
  {Kryder}}, \bibinfo {author} {\bibfnamefont {E.~C.}\ \bibnamefont {Gage}},
  \bibinfo {author} {\bibfnamefont {T.~W.}\ \bibnamefont {McDaniel}}, \bibinfo
  {author} {\bibfnamefont {W.~A.}\ \bibnamefont {Challener}}, \bibinfo {author}
  {\bibfnamefont {R.~E.}\ \bibnamefont {Rottmayer}}, \bibinfo {author}
  {\bibfnamefont {G.}~\bibnamefont {Ju}}, \bibinfo {author} {\bibfnamefont
  {Y.-T.}\ \bibnamefont {Hsia}}, \ and\ \bibinfo {author} {\bibfnamefont
  {M.~F.}\ \bibnamefont {Erden}},\ }\href {\doibase 10.1109/JPROC.2008.2004315}
  {\bibfield  {journal} {\bibinfo  {journal} {Proc. IEEE}\ }\textbf {\bibinfo
  {volume} {96}},\ \bibinfo {pages} {1810} (\bibinfo {year}
  {2008})}\BibitemShut {NoStop}%
\bibitem [{\citenamefont {Nowak}(2007)}]{Nowak2007}%
  \BibitemOpen
  \bibfield  {author} {\bibinfo {author} {\bibfnamefont {U.}~\bibnamefont
  {Nowak}},\ }\href
  {https://onlinelibrary.wiley.com/doi/abs/10.1002/9780470022184.hmm205}
  {\bibfield  {journal} {\bibinfo  {journal} {\textit{Classical Spin Models} in
  \textit{Handbook of Magnetism and Advanced Materials, vol. 2,
  Micromagnetism}}\ } (\bibinfo {year} {John Wiley \& Sons, 2007})}\BibitemShut
  {NoStop}%
\bibitem [{\citenamefont {Evans}\ \emph {et~al.}(2014)\citenamefont {Evans},
  \citenamefont {Fan}, \citenamefont {Chureemart}, \citenamefont {Ostler},
  \citenamefont {Ellis},\ and\ \citenamefont {Chantrell}}]{Evans2014}%
  \BibitemOpen
  \bibfield  {author} {\bibinfo {author} {\bibfnamefont {R.~F.~L.}\
  \bibnamefont {Evans}}, \bibinfo {author} {\bibfnamefont {W.~J.}\ \bibnamefont
  {Fan}}, \bibinfo {author} {\bibfnamefont {P.}~\bibnamefont {Chureemart}},
  \bibinfo {author} {\bibfnamefont {T.~A.}\ \bibnamefont {Ostler}}, \bibinfo
  {author} {\bibfnamefont {M.~O.~A.}\ \bibnamefont {Ellis}}, \ and\ \bibinfo
  {author} {\bibfnamefont {R.~W.}\ \bibnamefont {Chantrell}},\ }\href
  {https://iopscience.iop.org/article/10.1088/0953-8984/26/10/103202/meta}
  {\bibfield  {journal} {\bibinfo  {journal} {J. Phys. Condens. Matter}\
  }\textbf {\bibinfo {volume} {26}},\ \bibinfo {pages} {103202} (\bibinfo
  {year} {2014})}\BibitemShut {NoStop}%
\bibitem [{\citenamefont {Garanin}(1996)}]{Garanin1996}%
  \BibitemOpen
  \bibfield  {author} {\bibinfo {author} {\bibfnamefont {D.~A.}\ \bibnamefont
  {Garanin}},\ }\href {\doibase 10.1103/PhysRevB.53.11593} {\bibfield
  {journal} {\bibinfo  {journal} {Phys. Rev. B}\ }\textbf {\bibinfo {volume}
  {53}},\ \bibinfo {pages} {11593} (\bibinfo {year} {1996})}\BibitemShut
  {NoStop}%
\bibitem [{\citenamefont {Barthem}\ \emph {et~al.}(2013)\citenamefont
  {Barthem}, \citenamefont {Colin}, \citenamefont {Mayaffre}, \citenamefont
  {Julien},\ and\ \citenamefont {Givord}}]{barthem2013revealing}%
  \BibitemOpen
  \bibfield  {author} {\bibinfo {author} {\bibfnamefont {V.}~\bibnamefont
  {Barthem}}, \bibinfo {author} {\bibfnamefont {C.}~\bibnamefont {Colin}},
  \bibinfo {author} {\bibfnamefont {H.}~\bibnamefont {Mayaffre}}, \bibinfo
  {author} {\bibfnamefont {M.-H.}\ \bibnamefont {Julien}}, \ and\ \bibinfo
  {author} {\bibfnamefont {D.}~\bibnamefont {Givord}},\ }\href
  {https://www.nature.com/articles/ncomms3892} {\bibfield  {journal} {\bibinfo
  {journal} {Nat. Commun.}\ }\textbf {\bibinfo {volume} {4}},\ \bibinfo {pages}
  {2892} (\bibinfo {year} {2013})}\BibitemShut {NoStop}%
\bibitem [{\citenamefont {Wadley}\ \emph {et~al.}(2015)\citenamefont {Wadley},
  \citenamefont {Hills}, \citenamefont {Shahedkhah}, \citenamefont {Edmonds},
  \citenamefont {Campion}, \citenamefont {Nov{\'a}k}, \citenamefont
  {Ouladdiaf}, \citenamefont {Khalyavin}, \citenamefont {Langridge},
  \citenamefont {Saidl}, \citenamefont {Nemec}, \citenamefont {Rushforth},
  \citenamefont {Gallagher}, \citenamefont {Dhesi}, \citenamefont
  {Maccherozzi}, \citenamefont {{\v{Z}}elezn{\'y}},\ and\ \citenamefont
  {Jungwirth}}]{Wadley2015}%
  \BibitemOpen
  \bibfield  {author} {\bibinfo {author} {\bibfnamefont {P.}~\bibnamefont
  {Wadley}}, \bibinfo {author} {\bibfnamefont {V.}~\bibnamefont {Hills}},
  \bibinfo {author} {\bibfnamefont {M.~R.}\ \bibnamefont {Shahedkhah}},
  \bibinfo {author} {\bibfnamefont {K.~W.}\ \bibnamefont {Edmonds}}, \bibinfo
  {author} {\bibfnamefont {R.~P.}\ \bibnamefont {Campion}}, \bibinfo {author}
  {\bibfnamefont {V.}~\bibnamefont {Nov{\'a}k}}, \bibinfo {author}
  {\bibfnamefont {B.}~\bibnamefont {Ouladdiaf}}, \bibinfo {author}
  {\bibfnamefont {D.}~\bibnamefont {Khalyavin}}, \bibinfo {author}
  {\bibfnamefont {S.}~\bibnamefont {Langridge}}, \bibinfo {author}
  {\bibfnamefont {V.}~\bibnamefont {Saidl}}, \bibinfo {author} {\bibfnamefont
  {P.}~\bibnamefont {Nemec}}, \bibinfo {author} {\bibfnamefont {A.~W.}\
  \bibnamefont {Rushforth}}, \bibinfo {author} {\bibfnamefont {B.~L.}\
  \bibnamefont {Gallagher}}, \bibinfo {author} {\bibfnamefont {S.~S.}\
  \bibnamefont {Dhesi}}, \bibinfo {author} {\bibfnamefont {F.}~\bibnamefont
  {Maccherozzi}}, \bibinfo {author} {\bibfnamefont {J.}~\bibnamefont
  {{\v{Z}}elezn{\'y}}}, \ and\ \bibinfo {author} {\bibfnamefont
  {T.}~\bibnamefont {Jungwirth}},\ }\href {\doibase 10.1038/srep17079}
  {\bibfield  {journal} {\bibinfo  {journal} {Sci. Rep.}\ }\textbf {\bibinfo
  {volume} {5}},\ \bibinfo {pages} {17079} (\bibinfo {year}
  {2015})}\BibitemShut {NoStop}%
\bibitem [{\citenamefont {Atxitia}\ \emph {et~al.}(2010)\citenamefont
  {Atxitia}, \citenamefont {Hinzke}, \citenamefont {Chubykalo-Fesenko},
  \citenamefont {Nowak}, \citenamefont {Kachkachi}, \citenamefont {Mryasov},
  \citenamefont {Evans},\ and\ \citenamefont
  {Chantrell}}]{atxitia2010multiscale}%
  \BibitemOpen
  \bibfield  {author} {\bibinfo {author} {\bibfnamefont {U.}~\bibnamefont
  {Atxitia}}, \bibinfo {author} {\bibfnamefont {D.}~\bibnamefont {Hinzke}},
  \bibinfo {author} {\bibfnamefont {O.}~\bibnamefont {Chubykalo-Fesenko}},
  \bibinfo {author} {\bibfnamefont {U.}~\bibnamefont {Nowak}}, \bibinfo
  {author} {\bibfnamefont {H.}~\bibnamefont {Kachkachi}}, \bibinfo {author}
  {\bibfnamefont {O.~N.}\ \bibnamefont {Mryasov}}, \bibinfo {author}
  {\bibfnamefont {R.~F.~L.}\ \bibnamefont {Evans}}, \ and\ \bibinfo {author}
  {\bibfnamefont {R.~W.}\ \bibnamefont {Chantrell}},\ }\href
  {https://journals.aps.org/prb/abstract/10.1103/PhysRevB.82.134440} {\bibfield
   {journal} {\bibinfo  {journal} {Phys. Rev. B}\ }\textbf {\bibinfo {volume}
  {82}},\ \bibinfo {pages} {134440} (\bibinfo {year} {2010})}\BibitemShut
  {NoStop}%
\bibitem [{\citenamefont {Evans}\ \emph {et~al.}(2020)\citenamefont {Evans},
  \citenamefont {R\'ozsa}, \citenamefont {Jenkins},\ and\ \citenamefont
  {Atxitia}}]{Evans2020}%
  \BibitemOpen
  \bibfield  {author} {\bibinfo {author} {\bibfnamefont {R.~F.~L.}\
  \bibnamefont {Evans}}, \bibinfo {author} {\bibfnamefont {L.}~\bibnamefont
  {R\'ozsa}}, \bibinfo {author} {\bibfnamefont {S.}~\bibnamefont {Jenkins}}, \
  and\ \bibinfo {author} {\bibfnamefont {U.}~\bibnamefont {Atxitia}},\ }\href
  {\doibase 10.1103/PhysRevB.102.020412} {\bibfield  {journal} {\bibinfo
  {journal} {Phys. Rev. B}\ }\textbf {\bibinfo {volume} {102}},\ \bibinfo
  {pages} {020412} (\bibinfo {year} {2020})}\BibitemShut {NoStop}%
\bibitem [{\citenamefont {Evans}\ \emph {et~al.}(2015)\citenamefont {Evans},
  \citenamefont {Atxitia},\ and\ \citenamefont
  {Chantrell}}]{evans2015quantitative}%
  \BibitemOpen
  \bibfield  {author} {\bibinfo {author} {\bibfnamefont {R.~F.~L.}\
  \bibnamefont {Evans}}, \bibinfo {author} {\bibfnamefont {U.}~\bibnamefont
  {Atxitia}}, \ and\ \bibinfo {author} {\bibfnamefont {R.~W.}\ \bibnamefont
  {Chantrell}},\ }\href
  {https://journals.aps.org/prb/abstract/10.1103/PhysRevB.91.144425} {\bibfield
   {journal} {\bibinfo  {journal} {Phys. Rev. B}\ }\textbf {\bibinfo {volume}
  {91}},\ \bibinfo {pages} {144425} (\bibinfo {year} {2015})}\BibitemShut
  {NoStop}%
\bibitem [{\citenamefont {Callen}\ and\ \citenamefont
  {Callen}(1966)}]{callen1966present}%
  \BibitemOpen
  \bibfield  {author} {\bibinfo {author} {\bibfnamefont {H.~B.}\ \bibnamefont
  {Callen}}\ and\ \bibinfo {author} {\bibfnamefont {E.}~\bibnamefont
  {Callen}},\ }\href
  {https://www.sciencedirect.com/science/article/abs/pii/0022369766900126}
  {\bibfield  {journal} {\bibinfo  {journal} {J. Phys. Chem. Solids}\ }\textbf
  {\bibinfo {volume} {27}},\ \bibinfo {pages} {1271} (\bibinfo {year}
  {1966})}\BibitemShut {NoStop}%
\bibitem [{\citenamefont {Garanin}(1997)}]{Garanin1997}%
  \BibitemOpen
  \bibfield  {author} {\bibinfo {author} {\bibfnamefont {D.~A.}\ \bibnamefont
  {Garanin}},\ }\href
  {https://journals.aps.org/prb/abstract/10.1103/PhysRevB.55.3050} {\bibfield
  {journal} {\bibinfo  {journal} {Phys. Rev. B}\ }\textbf {\bibinfo {volume}
  {55}},\ \bibinfo {pages} {3050} (\bibinfo {year} {1997})}\BibitemShut
  {NoStop}%
\bibitem [{\citenamefont {Garanin}\ and\ \citenamefont
  {Chubykalo-Fesenko}(2004)}]{Garanin2004}%
  \BibitemOpen
  \bibfield  {author} {\bibinfo {author} {\bibfnamefont {D.~A.}\ \bibnamefont
  {Garanin}}\ and\ \bibinfo {author} {\bibfnamefont {O.}~\bibnamefont
  {Chubykalo-Fesenko}},\ }\href
  {https://journals.aps.org/prb/abstract/10.1103/PhysRevB.70.212409} {\bibfield
   {journal} {\bibinfo  {journal} {Phys. Rev. B}\ }\textbf {\bibinfo {volume}
  {70}},\ \bibinfo {pages} {212409} (\bibinfo {year} {2004})}\BibitemShut
  {NoStop}%
\bibitem [{\citenamefont {Atxitia}\ \emph {et~al.}(2009)\citenamefont
  {Atxitia}, \citenamefont {Chubykalo-Fesenko}, \citenamefont {Chantrell},
  \citenamefont {Nowak},\ and\ \citenamefont {Rebei}}]{Atxitia2009}%
  \BibitemOpen
  \bibfield  {author} {\bibinfo {author} {\bibfnamefont {U.}~\bibnamefont
  {Atxitia}}, \bibinfo {author} {\bibfnamefont {O.}~\bibnamefont
  {Chubykalo-Fesenko}}, \bibinfo {author} {\bibfnamefont {R.~W.}\ \bibnamefont
  {Chantrell}}, \bibinfo {author} {\bibfnamefont {U.}~\bibnamefont {Nowak}}, \
  and\ \bibinfo {author} {\bibfnamefont {A.}~\bibnamefont {Rebei}},\ }\href
  {https://journals.aps.org/prl/abstract/10.1103/PhysRevLett.102.057203}
  {\bibfield  {journal} {\bibinfo  {journal} {Phys. Rev. Lett.}\ }\textbf
  {\bibinfo {volume} {102}},\ \bibinfo {pages} {057203} (\bibinfo {year}
  {2009})}\BibitemShut {NoStop}%
\bibitem [{\citenamefont {Jhuria}\ \emph {et~al.}(2020)\citenamefont {Jhuria},
  \citenamefont {Hohlfeld}, \citenamefont {Pattabi}, \citenamefont {Martin},
  \citenamefont {Arriola~Cordova}, \citenamefont {Shi}, \citenamefont
  {Lo~Conte}, \citenamefont {Petit-Watelot}, \citenamefont {Rojas-S\'anchez},
  \citenamefont {Malinowski} \emph {et~al.}}]{jhuria2020spin}%
  \BibitemOpen
  \bibfield  {author} {\bibinfo {author} {\bibfnamefont {K.}~\bibnamefont
  {Jhuria}}, \bibinfo {author} {\bibfnamefont {J.}~\bibnamefont {Hohlfeld}},
  \bibinfo {author} {\bibfnamefont {A.}~\bibnamefont {Pattabi}}, \bibinfo
  {author} {\bibfnamefont {E.}~\bibnamefont {Martin}}, \bibinfo {author}
  {\bibfnamefont {A.~Y.}\ \bibnamefont {Arriola~Cordova}}, \bibinfo {author}
  {\bibfnamefont {X.}~\bibnamefont {Shi}}, \bibinfo {author} {\bibfnamefont
  {R.}~\bibnamefont {Lo~Conte}}, \bibinfo {author} {\bibfnamefont
  {S.}~\bibnamefont {Petit-Watelot}}, \bibinfo {author} {\bibfnamefont {J.~C.}\
  \bibnamefont {Rojas-S\'anchez}}, \bibinfo {author} {\bibfnamefont
  {G.}~\bibnamefont {Malinowski}},  \emph {et~al.},\ }\href
  {https://www.nature.com/articles/s41928-020-00488-3} {\bibfield  {journal}
  {\bibinfo  {journal} {Nat. Electron.}\ }\textbf {\bibinfo {volume} {3}},\
  \bibinfo {pages} {680} (\bibinfo {year} {2020})}\BibitemShut {NoStop}%
\bibitem [{\citenamefont {Atxitia}\ \emph {et~al.}(2016)\citenamefont
  {Atxitia}, \citenamefont {Hinzke},\ and\ \citenamefont
  {Nowak}}]{atxitia2016fundamentals}%
  \BibitemOpen
  \bibfield  {author} {\bibinfo {author} {\bibfnamefont {U.}~\bibnamefont
  {Atxitia}}, \bibinfo {author} {\bibfnamefont {D.}~\bibnamefont {Hinzke}}, \
  and\ \bibinfo {author} {\bibfnamefont {U.}~\bibnamefont {Nowak}},\ }\href
  {https://iopscience.iop.org/article/10.1088/1361-6463/50/3/033003/meta?casa_token=9ij2ilcQ7IQAAAAA:1pGQjN7mYXUecmUBJgHj5Wjmh3nICj4zceUVFAzFizCnoOXLrDKOooF103vlwemRoMlqWNLC}
  {\bibfield  {journal} {\bibinfo  {journal} {J. Phys. D: Appl. Phys.}\
  }\textbf {\bibinfo {volume} {50}},\ \bibinfo {pages} {033003} (\bibinfo
  {year} {2016})}\BibitemShut {NoStop}%
\bibitem [{\citenamefont {Hirst}\ \emph {et~al.}(2022)\citenamefont {Hirst},
  \citenamefont {Atxitia}, \citenamefont {Ruta}, \citenamefont {Jackson},
  \citenamefont {Petit},\ and\ \citenamefont {Ostler}}]{hirst2022multiscale}%
  \BibitemOpen
  \bibfield  {author} {\bibinfo {author} {\bibfnamefont {J.}~\bibnamefont
  {Hirst}}, \bibinfo {author} {\bibfnamefont {U.}~\bibnamefont {Atxitia}},
  \bibinfo {author} {\bibfnamefont {S.}~\bibnamefont {Ruta}}, \bibinfo {author}
  {\bibfnamefont {J.}~\bibnamefont {Jackson}}, \bibinfo {author} {\bibfnamefont
  {L.}~\bibnamefont {Petit}}, \ and\ \bibinfo {author} {\bibfnamefont
  {T.}~\bibnamefont {Ostler}},\ }\href {https://arxiv.org/abs/2206.08625}
  {\bibfield  {journal} {\bibinfo  {journal} {arXiv preprint arXiv:2206.08625}\
  } (\bibinfo {year} {2022})}\BibitemShut {NoStop}%
\end{thebibliography}%

\end{document}